
\magnification=1200
\settabs 18 \columns

\baselineskip=12pt

\font\steptwo=cmb10 scaled\magstep2
\font\stepthree=cmb10 scaled\magstep3

\def\b{\bigskip}
\def\bb{\bigskip\bigskip}

\def\sqr#1#2{{\vcenter{\vbox{\hrule height.#2pt
 \hbox{\vrule width.#2pt height#1pt \kern#1pt
 \vrule width.#2pt} \hrule height.#2pt}}}}
\def\square{\mathchoice \sqr65 \sqr65 \sqr{2.1}3 \sqr{1.5}3}

\def\operp{\hbox{${\kern+.25em{\bigcirc}
\kern-.85em\bot\kern+.85em\kern-.25em}$}}

\def\lsim{\;\raise0.3ex\hbox{$<$\kern-0.75em\raise-1.1ex\hbox{$\sim$}}\;}
\def\gsim{\;\raise0.3ex\hbox{$>$\kern-0.75em\raise-1.1ex\hbox{$\sim$}}\;}
\def\no{\noindent}
\def\r{\rightline}
\def\ce{\centerline}
\def\ve{\vfill\eject}
\def\rdots{\mathinner{\mkern1mu\raise1pt\vbox{\kern7pt\hbox{.}}\mkern2mu
 \raise4pt\hbox{.}\mkern2mu\raise7pt\hbox{.}\mkern1mu}}

\def\e e{$e^+ e^-$ }

\pageno=1
\r {hep-th/9806072}
\r {CERN-TH/98-186}
\r {UCLA/98/TEP/17}
\def\today{\ifcase\month\or January\or February\or
March\or April\or  May\or June\or July\or August\or
September\or October\or November\or  December\fi
\space\number\day, \number\year }


\vskip1.5in
\ce{\stepthree GAUGE FIELDS AND SINGLETONS OF AdS$_{\hbox {\steptwo 2p+1}}$}
\vskip1.0cm
\ce{Sergio Ferrara* and Christian Fronsdal**}
\b
\ce{\it *CERN, Geneva, Switzerland.}
\ce{\it **Physics Department, University of California, Los Angeles, CA
90090-1547.}
\vskip3.0cm

\no {\it ABSTRACT.}  We show that $p$-forms on  $AdS_{2p+1}$ describe both
singletons and
massless particles. On the $2p$-dimensional boundary the singleton $p$-form
Lagrangian
reduces to the conformally invariant functional $\int F^2$. All the
representations,
singletons as well as massless, are zero center modules and involve a
vacuum mode.   Two- and
three-form singleton fields are required by supersymmetry in AdS$_5$ and
AdS$_7$ supergravity
respectively.

\vskip 6truecm
 \today
\ve
\voffset0pt

\no{\steptwo 1. Introduction}

\parskip.1cm

It is well known that spin-0 singletons on $AdS_{d+1}$ can be extended to
``super singletons" by introducing spin-1/2 particles [1].  This fact can be
related to the property of the massless Klein-Gordon and Dirac equations,
on the boundary at infinity, of being
conformally invariant.
In this note we point out that whenever the space-time dimension $d$ is even,
  Maxwell's  equations for the $p = d/2$-form field strength are also
conformally invariant, and therefore one expects
$(p-1)$-form potentials to describe degrees of freedom of the singleton
type on $AdS_{d+1}$. [2],[3],[4].

This is a generalization of what is already known for $AdS_5$ and
$AdS_7$ in which case supersymmetry does indeed require one- [3] and two-form
gauge potentials respectively [4].

The extension of $2p$-dimensional Maxwell theory to a singleton theory in
$AdS_{2p+1}$ is a generalization of
the circumstance that superconformal invariance at the boundary of $AdS_5$
and $AdS_7$ requires the singletons to be
accompanied by 1-form and 2-form potentials, respectively, in the same
super multiplet.
The extension of 4-dimensional Maxwell theory to a singleton theory on
$AdS_5$ was described in two papers by the same authors
[5],[6].  There it was pointed out that the Maxwell field in four
dimensions transforms as zero-center module of
$SO(4,2)$. Other important examples of zero-center modules are in [7].

In terms of $SO(4,2)$ irreducibles, as usual denoted $D(E_0,J_1,J_2)$, the
complete non-decomposable representation
carried by the Maxwell potential in four dimensions is [8]
$$
D(1, \raise.5mm\hbox{$ \scriptstyle {1\over 2}$},\raise.5mm\hbox{$
\scriptstyle {1\over 2}$})
\rightarrow \bigl(D(2,1,0) \oplus D(2,0,1) + {\rm id}\bigr) \rightarrow
D(1, \raise.5mm\hbox{$ \scriptstyle {1\over 2}$},\raise.5mm\hbox{$
\scriptstyle {1\over 2}$}),
$$
where $D(2,1,0)$ is associated with the self-dual part of the field
strength. The quantum numbers $E_0,J_1,J_2$ refer to
the highest weight, a weight of $O(2) \otimes SO(4)$.

In the six-dimensional conformal field theory the field representations can
be denoted $D(E_0,a_1,a_2,a_3)$, where $a_1,a_2,a_3$
are $SO(6)$ Dynkin labels. The three-form (self-dual) field strength
carries $D(3,2,0,0)$. The nondecomposable representation
is in this case
$$
D(2,1,0,1) \rightarrow \bigl( D(3,2,0,0) \oplus D(3,0,0,2) \oplus
D(1,0,1,0)\bigr) \rightarrow D(2,1,0,1).
$$
This is again a zero-center module, of $SO(6,2)$. The representations that
are carried by the extensions of these boundary
conformal fields to $AdS_{d+1}$ are more complicated, though the central,
physical subquotients are the same.
The non-decomposable character of the representation is of course the
characteristic property of gauge theories.

It is worth while to notice that these are precisely the forms that can be
``self-dual",
in any even-dimensional space,\footnote *{Self-dual ``real" $p$-forms exist
for $p = 4k+2$, for
$p = 4k$ the forms must be complexified.} and that can be coupled to
$(p-1)$-brane dyons. This leads to the following conjecture, that will be
proved in this paper. Conformal field strengths
of degree $p = d/2$ on the boundary of $AdS_{d+1}$ are zero-center modules
for any even $d$.

Actually, the hierarchy of $q$-forms that carry zero-center $SO(2p,2)$
modules goes from $q = 0$ to $q = p$, with conformal degree
$q$ for singletons and conformal degree $p+q$ for representations that are
associated with ``massless" field theories in the bulk.
The space of $(p-1)$-form singleton potential field modes carries
\underbar{all} of these representations. The
scalar field (conformal dimension $2p$) and  the vector potential
(conformal dimension $2p-1$) are of the type that occur in
supergravities, the others seem to belong to the same category and they
will therefore be referred to as ``massless" in this
paper. These forms, of rank higher than 1, do not appear in spectra of
higher dimensional supergravities; they represent something
that is new and perhaps interesting.

$$
\hskip-1.8cm E_0\uparrow
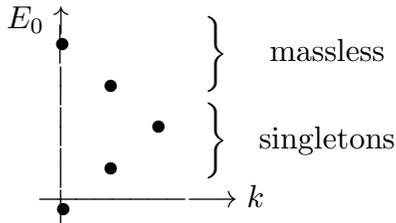
$$
\vskip-.7cm
$$
\eqalign{
& \hskip-2mm\matrix{\big|\hskip-1.2mm\bullet \cr \hskip-.3mm\big|&\bullet
\cr \hskip-.4mm\big|&&\bullet \cr \hskip-.3mm\bigl|&
\bullet
\cr
\hskip.2mm\big|\hskip-1.2mm\bullet
\cr}
\cr}
$$
\vskip-.9cm
$$
\hskip.7cm-\hskip-1.5mm -\hskip-1.5mm -\hskip-1.5mm -\hskip-1.5mm
-\hskip-1.5mm -\hskip-1.5mm -\hskip-1.5mm -\hskip-1.mm
-\hskip-1.5mm
\rightarrow k
$$
\vskip-2.8cm
$$
\hskip4cm \matrix{\biggl\}&{\rm massless}\cr  \cr\biggr\}&{\rm singletons}\cr}
$$
\b
In this diagram, drawn for the case $p = 2$, we show the zero-center
modules \break
$D(p\pm k,w_{p-k})$ ($w_k$ is a highest weight of $SO(2p)$) as dots at the
points $E_0 = p\pm k$. We show that there are extensions
between neighbour  representations, for all $p$.

In Section 2 we study gauge theories of the conventional type on
$AdS_{2p+1}$. These potentials are $(p-1)$-forms of conformal
degree $p+1$, and we call them massless.  They are massless
merely in the sense that they have a familiar gauge structure, but
\underbar {not} in the sense that they appear in supergravities.
Instead, what appear in supergravities (in the bulk) are certain
$(p-1)$-forms (with $E_0 = 2p-1)$ that satisfy self-duality
constraints  of the  type $mA = ^*dA$ [4]. These forms occur for $p=2,3$ in
$AdS_{2p+1}$ supergravity and can be thought of as
``composite" boundary operators of a scalar singleton and a $p$-form singleton field strength.

In Section 3 we concentrate on
$(p-1)$-form potentials of conformal degree
$p-1$ and show that they have gauge sectors of the singleton type (and of
conventional type as well).  All these representations
are  zero-center modules. What is even more remarkable is that the massless
representations appear in the gauge sector of the
singleton representations. Section 4 summarizes the conclusions and the
Appendix gives an alternative derivation of some of the
results.

\bb
\no{\steptwo 2. Spaces, Groups and Highest Weight representations.}

\no {\it2.1.}  The space $M=AdS_{2p+1}$ is a space-time with one time and
$2p$ space
dimensions, endowed with an anti-De Sitter metric.  The conformal completion
$\bar M$ of this space includes a boundary $M_\infty$ with a Minkowski metric,
signature $1,2p-1$.

This paper is a study of a field theory on $M_\infty$, the principal field
a $(p-1)$-form $A_\infty$ (the potential), with associated field strength
$F_\infty = dA_\infty$ and action $\int F_\infty^2$, and mainly the
extension of this theory to a singleton topological gauge theory on $\bar M$.
We suppose that $F_\infty$ has conformal dimension $p$, making the boundary
action
conformally invariant.

Let $r$ be a radial coordinate on $\bar M$.  The extension $A$ of
$A_\infty$ to $\bar M$ is a $(p-1)$-form on $\bar M$ that is related to
$A_\infty$ by [9][5]
$$
\lim_{r\to\infty} r^NA = A_\infty~,\eqno(2.1)
$$
for some real number $N$.  This boundary condition on $A$ determines the
conformal degree of $A_\infty$; it is $-N$.  Hence $N=1-p$.  (Of course,
the identification (2.1) implies a projection of the left-hand side from
the tangent space of $\bar M$ to the tangent space of $M_\infty$.)

We shall show that the field $A$ on $\bar M$, $M=AdS_{2p+1}$, is a gauge
theory, in two ways.
The logic of the investigation leads us to include a study of $p$-forms of
conformal dimension
$p+2$ as well. These fields describe massless particles on AdS$_{2p+1}$.
\bb

\no{\it 2.2. Notation}.

The space $AdS_{2p+1}$ is the homogeneous space $SO(2p,2)/SO(2p,1)$.  For
calculational purposes we represent it as the hyperboloid
$$
y^2=1~, \quad
y^2 := y_0^2 + y_{00}^2 - \sum^{2p}_{i=1} y_i^2~. \eqno(2.2)
$$
A considerable simplification results from extending our fields on $M$ to
the domain
$$
M_+ = \{y \in {\bf R}^{2p+2},~y^2>0\}~.
$$
This is done by fixing the degree of homogeneity of the extended fields.
Thus $A$ becomes a homogeneous field on $M_+$.  In fact, we extend $A$ to a
homogeneous
$(p-1)$-form on $M_+$. This can be done by  stipulating that this
$(p-1)$-form be transverse,
or otherwise as the situation favors it.  We continue to use the same
letter $A$ for this
extended field. If the degree of homogeneity of $A$ coincides with the
number $N$ in (2.1),
then $A_\infty = A|_{y^2 = 0}$. The boundary cone $y^2 = 0$ of $M_+$ can
thus be identified with the boundary $M_\infty$ of $M$
[10].

The principal advantage of this trick is that the vector fields
$\partial_\alpha = \partial/\partial y^\alpha,~\alpha=0,00,1,\ldots,2p$, are
now well defined.  The degree $N$ of $A$ is given by
$$
y^\alpha\partial_\alpha A = y\cdot\partial_y A = NA~. \eqno(2.3)
$$
For the time being we look at $N$ as a free parameter that we can choose to
suit our convenience.
\b

\no{\it 2.3. Tensor Gauge Theory.}

We try to impose conditions on the $(p-1)$-form $A$ so as to make it the
carrier of an
irreducible representation of $SO(2p,2)$:
$$
\eqalignno{&y\cdot A = 0~, & (2.4) \cr
&(y^2\partial^2 + \kappa)A = 0~, & (2.5) \cr
&\partial\cdot A = 0~, & (2.6) \cr
&^*(dA) = \pm dA~. & (2.7) \cr}
$$
\no {\it Explanation}.  The vector field $y\cdot\partial$ is invariant,
Eq. (2.4) makes $A$ transverse to this radial field and assures its
interpretation as a $(p-1)$-form on $M$.  Because of (2.3), Eq. (2.5) is
equivalent to the covariant wave equation on $M$. We have
$$
\square = y^2\partial^2 - N(N+2p)~, \quad \partial^2 = \partial_\alpha
\partial^\alpha~. \eqno(2.8)
$$
Hence (2.5) is the same as
$$
(\square + m^2)A=0,~~m^2 :=\kappa - N(N+2p)~. \eqno(2.9)
$$
We choose $N$ so as to make $\kappa=0$.  Finally, in (2.7), $^*(dA)$ is
the Hodge dual.  Since we shall have no use for it we skip the
definition.

The representation of $SO(2p,2)$ that is induced on the space of solutions
of all these equations is not reducible, but it may admit an invariant
subspace of gauge fields of the form
$$
A = y^2d\Phi + ay\wedge\Phi~, \eqno(2.10)
$$
$\Phi$ a $(p-2)$-form.  A straightforward calculation gives the result that
this happens if and only if
$$
N+p = \pm 1~, \quad a=3-N-p~. \eqno(2.11)
$$
The appearance of this invariant subspace is characteristic of massless
representations.
There is also, when $N = 1-p$, an invariant subspace of field modes of the
form $d\Phi$, \footnote *{ We recall
that the boundary $M_\infty$ of $M$ can be identified with the boundary
cone $y^2 = 0$ of $M_+$. Since the fields are required to be
defined on $\bar M$, no inverse powers of $y^2$ are permitted in (2.10).
This why this case could not be included in (2.10).}
as well as an invariant subspaces of the form $y^2\Phi$; this last is what
characterizes singletons. For a more systematic
approach, that encompasses all the possibilities at once, see Appendix.
\bb

\no{\it 2.4.Highest Weight Representations, Massless Representations.}

The second-order Casimir operator ${\cal C}$ of $SO(2p,2)$ takes the form
$$
{\cal{C}}A = \bigl(-y^2\partial^2 + N(N+2p) + p^2-1\bigr)A, \eqno(2.12)
$$
\no when $A$ is constrained by (2.3), (2.4) and (2.6).  Using (2.5) we
obtain
$$
{\cal{C}}A = [(N+p)^2-1] A~,
$$
\no and when we require the existence of gauge modes, hence Eq. (2.11),
$${\cal{C}}A=0.$$

On a highest weight $D(E_0, w)$, where $E_0$ is the energy $L_{05}$ and
$ w$ is the highest weight of a representation of $SO(2p)$, ${\cal{C}}$
takes the value
$$
{\cal{C}}\to E_0(E_0-2p) + c(w), \eqno(2.13)
$$
\no where $c(w)$ is the value of the second-order Casimir operator of
$SO(2p)$ on the representation with highest weight $  w$.  If this
representation is the $k^{\rm th}$ alternating power of the fundamental
representation (real representation of dimension $2p$) then it is
$$
c( w_k) = k(2p-k)~. \eqno(2.14)
$$

The following set of representations,
$$
\eqalignno{&D(k,w_k),~~ k=0,1,\cdots,p~, & (2.15) \cr
&D(p+k,w_{p-k}),~~ k=1,\ldots,p~, & (2.16) \cr}
$$
\no all have ${\cal{C}}=0$.  In fact, they are all zero-center modules;
that is, all the
Casimirs operators are zero on these representations. The proof is in the
Appendix.

We shall show here the existence of an extension, within the field module,
of the type
$$
D(p+1,w_{p-1})\to D(p+2,w_{p-2})~. \eqno(2.17)
$$
Let $z=\{z_\alpha\}_{\alpha=0,00,1,\ldots,2p}$ be an auxiliary Grassmannian
vector ``coordinate", and represent $k$-forms on $TM_+$ as polynomials of order
$k$ in $z$.  For example, the gauge field (2.10) beomes
$$
y^2z\cdot\partial_y\Phi+ay\cdot z\Phi~. \eqno(2.18)
$$
A very simple ground state with $N=-p-1$ is
$$
y_+^{-2-p}(y_+-z_+y\cdot\partial_z)\Psi~, \eqno(2.19)
$$
\no with $\Psi$ a polynomial of order $p-1$ in $z_1,\ldots,z_{2p}$ only,
and
$$
y_+ = y_0+iy_{00} \propto re^{it}~, \quad
z_+ = z_0+iz_{00}~.
$$
\no It is a highest weight vector for the representation $D(p+1,w_{p-1})$,
where $w_{p-1}$ is the highest weight of the $(p-1)^{\rm th}$ exterior
power of the fundamental representation $D_1$ of $SO(2p)$.  According to (2.13)
and (2.14), the Casimir takes the value 0.

We shall now show that the space generated from (2.19) (the Harish-Chandra
module) contains an invariant subspace that is generated from a singular
vector of the form (2.18).

Let $\{L_1^i\}_{i=1,\ldots,2p}$ be a set of step-up operators for the energy;
they span a space of dimension $2p$ that, under the restriction of the
adjoint action to $SO(2p)$, carries the fundamental representation $D_1$ of
$SO(2p)$.  Applying these operators to (2.19) we obtain an $SO(2p)~
p$-form that can
be contracted to a $(p-2)$-form with the $SO(2p)$ metric, which gives us a
vector with weight $(p+2,w_{p-2})$.  The representation $D(p+2,w_{p-2})$ has
${\cal{C}}=0$ (see above) and this state is thus a candidate for being the
highest weight vector of an invariant submodule.

For completeness, we present the details of the calculation. The energy
raising operators are
$$
L_+^i = y_i\partial_+ + y_-\partial_i + z_i\partial_+^\prime +
z_-\partial_i^\prime, ~~i=1,\ldots,2p~.
$$
\no The following function is a highest weight vector with weight
$(p+1,w_{p-1})$,
$$
y_+^{-p-2}A_{i_1\ldots i_{p-2}},~~ A_{i_1\ldots i_k}= (y_+ - z_+\vec
y\cdot\vec\partial_z)z_{i_1}
\ldots z_{i_k}~. \eqno(2.20)
$$
\no We find
$$
L_+^i \bigl[y_+^{-p-2}
A_{ii_1\ldots i_{p-2}}\bigr] = {p+2\over 4} (y^2z\cdot \partial_y+4y\cdot z)
\bigl[y_+^{-p-3}A_{i_1\ldots i_{p-2}}\bigr]~.
$$
\no This shows that the extension (2.17) actually occurs in the space
generated by the ground state (2.20), among the modes of a $(p-1)$-form
potential.
\bb

\no{\it 2.5. Highest Weight Representations: Singletons.}

We turn to an analysis of the case that is of main interest to us, the case
$N+p=1$ in (2.11), so we are now dealing with $(p-1)$ forms of degree $p-1$
and field strengths of degree $p$, as required by the conformal invariance
of $\int F^2$ on the $2p$-dimensional boundary.

This case is quite different, for an extension of the type (2.17),
linking $D(p-1,w_{p-1})$ to $D(p,w_{p-2})$, is not possible, the two
representations having different values of ${\cal{C}}$.

Instead, there is the extension
$$
D(p,w_p)\to D(p-1,w_{p-1})~. \eqno(2.21)
$$
The ``ground state" of $D(p,w_p)$ is
$$
y_+^{-p} y\wedge \psi~, \quad
\psi_{i_1\ldots i_{p-1}} = z_{i_1} \ldots z_{i_{p-1}}~. \eqno(2.22)
$$
\no Applying the lowering operators we find
$$
L_-^i(y_+^{-p}y\wedge\psi)_{ii_p\ldots i_{p-1}} \propto
y_+^{-p}(y_+-z_+\vec y\cdot\vec\partial_z)\psi_{i_1\ldots i_{p-1}}~.
\eqno(2.23)
$$
\no This is the ground state of $D(p-1,w_{p-1})$.  Pushing up again, one
obtains a mode with
symmetry different from that of (2.22), orthogonal to it. Hence (2.22) is a
ground state modulo the
invariant subspace generated from (2.23).

This case is indeed very different.  There is a gauge subspace generated by
(2.23).  The cyclic vector for
(2.21) is (2.22), which is not the highest weight, so (2.21) is not a
Harish-Chandra module.

The extension (2.21) does not reveal the full extent of the field multiplet
carried by the $(p-1)$-form quantum field operator.  We shall show, in
Section 3.2,  that there are
extensions to the massless representation $D(p+1,w_{p-1})$ and to
$D(p-2,w_{p-2})$.

\bb

\no{\steptwo 3. The Quantum Field Operator.}

The degree of homogeneity of the $(p-1)$-form $A$ can be chosen so that each
component satisfies the scalar wave equation
$$
\partial^2A = 0 \quad \hbox{or} \quad \partial^2A_{\alpha_1\ldots\alpha_{p-1}}
=0~.
$$
\no Choose a space of solutions of the scalar wave equation, and let
$D_0$ be the highest weight representation of $SO(2p,2)$ induced on this
space.  Then the representation carried by the field $A$ is a subquotient
of the direct product
$$D_0 \otimes D_{p-1}~, \quad D_{p-1} = D_1^{\wedge (p-1)},
$$
\no where $D_1$ is the fundamental (vector) representation of
$SO(2p,2)$.  Now suppose that $D_0 = D(E_0,0)$, then for $E_0$ big
enough, this product is completely reducible.  But if $E_0=p\pm 1$, then
this is not the case.

The strategy that will uncover all possibilities is thus to calculate the
structure of the representations
$$
D(p\pm 1,0) \otimes D_{p-1}~.
$$
But first we need to recall a special property of the representation
$D(p-1,0)$.

There is a scalar field that carries the representation
$D(p+1,0)$, but there is no scalar field that carries just
$D(p-1,0)$.  This phenomenon has already been explained several times,
in dimensions 3, 4, 5 and even in the general case. See [10], and [2] for a
review. Therefore, let us just
recall the fact that the modes of $D(p-1,0)$ appear in the scalar
quantum field operator inside a complete Gupta-Bleuler triplet, namely
$$
D(p+1, 0) \to D(p-1, 0) \to D(p+1,0) =: D_S~.\eqno(3.1)
$$
\no The ground state for this representation  is the highest weight of
$D(p-1,0)$, namely\footnote *
{The ground state of the left factor of (3.1), which is cyclic for the
whole representation, is
$py^2y_+^{-p-1}\,$ln$\,y_+ + 2y_+^{-p}y_-$.}
$$
y_+^{1-p}.
$$
It is easy to show that
$$
\sum_i (L_+^i)^2 y_+^{1-p} ~\propto ~y^2y_+^{-1-p}~,
$$
\no hence $D(p+1,0)$ appears as a subrepresentation of the
Harish-Chandra representation generated from the ground state of
$D(p-1,0)$.  Consequently, we shall have to investigate
$$
D(p+1, 0) \otimes D_{p-1}, ~{\rm (massless ~fields)} \quad \hbox{and} \quad
D_S \otimes D_{p-1}~{\rm (singletons)}.
$$
\no The first case is a type of generalized electrodynamics (in the bulk).
The second
case leads to tensor singleton fields; they are gauge fields both in the
traditional sense and (in the bulk) in the topological singleton sense.

The massless case is quite simple: we give the result without proof:
$$
D(p+1,0) \otimes D_{p-1} = D(p+2,0) \oplus D(p,0) \oplus D_{\rm massless}
$$
(omit the first term when $p = 2$), with
$$
D_{\rm massless} = D(p+2,w_{p-2}) \rightarrow D(p+1,w_{p-1}) \rightarrow
D(p+2,w_{p-2}).\eqno(3.2)
$$
The direct summands can be eliminated by requiring the field operator to
satisfy
the constraint $(y^2\partial_y\cdot \partial_z + 2y\cdot \partial_z)A = 0$.
The Lorentz condition is $y\cdot A = 0$, and the invariant subspace of
gauge modes consists of
potentials of the form $(y^2z\cdot \partial_y + 4y\cdot z)\Phi$.\footnote
*{The ground states of the
three subquotients, reading from right to left,
are as follows, $ y_+^{-3}z_+,~ y_+^{-3}z_- - y_+^{-4}\vec y\cdot \vec z +
(y_+^{-5}\vec y\,^2 -
y_+^{-4}y_-)z_+$ and
$2y_+^{-3}z_- - y_+^{-4}\vec y\cdot \vec z$.}\footnote {**}{ This is not
the only way, and perhaps not the best way.}

The propagators of the massless field that carries the entire product
representation $D(p+1,0) \otimes D_1^{\wedge(p+1)}$ is
$$
(z\cdot z')^{p+1}K_1(y\cdot y'),~~ K(y\cdot y') \propto {}_2F_3(
$$
There is another propagator, for the transverse vector field, that
carries the representation (3.2) only.\footnote {**}{Give it, and the
hypergeometric above and below.}

\b
\no {\it 3.2 Singleton gauge theory, the field module.}
This is a single Gupta-Bleuler triplet,
$$
D_S\otimes D_1 = D_{\rm scalar}\to D_{\rm physical}\to D_{\rm gauge},
\eqno(3.3)
$$
\no with the two outer terms equivalent.  No covariant constraint can be
imposed on the quantum field operator and the propagator is simply
$(z\cdot z^\prime)^{p-1} K(y,y^\prime)$ where $K$  is the propagator for
$D_S$, see below.
The gauge sector is very
complicated.  The Lorentz condition projects the theory on the boundary,
where it is a conformal formulation of ordinary electrodynamics---Maxwell
theory to be precise.  The best way to formulate the theory is in terms of a
dipole field [11] that satisfies $(\square^2+m^2)^2A = 0$, or
$$
(\square + m^2)A=B~, \quad
(\square + m^2)B=0~.
$$
\no Then the Lorentz condition includes the constraints $B=0$ and
$y\cdot A = 0$.

We shall now justify these remarks, and at the same time demonstrate that
the field module includes an
extension of the singleton representation to the massless representation;
that is, we show that
the extension
$$
D(p,w_p) \rightarrow D(p+1,w_{p-1})
$$
actually occurs. The calculation is presented here for the case $p = 2$ only.

 The ground state $y_+^{-2}(y \wedge z)$ of $D(2,w_2)$ is a relative one
that pushes down to
the gauge mode $y_+^{-2}(y_+\vec z - \vec y z_+) = z\cdot \partial_y
(y_+^{-1}\vec y)$. Examination
of the propagator at levels 0 and 1 shows that the conjugate of this mode
is $y_+^{-1}\vec z$, modulo
gauge. This one pushes down to the vacuum mode $z_+/y_+$, which shows that
the latter is part of the
physical sector. It also pushes up, and at level 3 we find in particular
the mode
$$
L_{+i}L_+^j(y_+^{-1}z_j) - L_+^jL_{+j}(y_+^{-1}z_i) = 6y^2y_+^{-4}(y_+ z_i
- y_iz_+),
$$
which is the ground state of $D(3,w_1)$. The complete reduction of the
singleton field module is thus
$$
D_S \oplus D_1 = D(1,w_1) \rightarrow \big[D_{\rm massless} \oplus D(2,0)
\oplus {\rm id}\bigr] \rightarrow D(1,w_1).
$$
The singleton propagator is
$$
(z\cdot z')^{p-1}{}_3F_2(
$$
The hypergeometric function is of logarithmic type.

\bb

\no{\steptwo 4. Conclusions.}

In this paper we have shown that conformally invariant $p$-form field
strengths, of conformal degree $p$, and the associated
$(p-1)$-form potentials, in
$2p$-dimensional Minkowski space, can be extended to a topological
singleton field theory in $AdS_{2p+1}$. The representations of
$SO(2p,2)$ carried by these fields are all zero-center modules, just as is
the case for conformal electrodynamics in $4$ dimensions
(the case $p = 2$).

Field strengths of conformal degree $p+2$ describe massless fields in the
bulk; they too are zero-center modules. In
fact, these representations appear as part of the gauge sector in the
singleton field module.

Fields of this kind are known to be required by superconformal symmetry
[6],[2],[3] in the case $p = 2,3$; they may play a role for
$p = 4,5$ as well, though  simple super extensions of $AdS_9$ and
$AdS_{11}$ do not exist. One may point out that the self-dual
five form field strength of the $10d,IIB$ string is a "singleton" 
candidate
for $AdS_{11}$ in the sense of this paper. This may have some implications
for
attempts to construct a fundamental framework for strings or M-theory.

\bb
\no{\steptwo Appendix.}

Here it will be shown that the following highest weight representations of
$SO(2p,2)$,
$$
D(k,w_k),~~ k = 0,...,p,
$$
are zero-center modules; that is, the center of the enveloping algebra is
represented by zero. The demonstration consists of
exhibiting equivariant maps from $D(k,w_k)$ to $D(k+1,w_{k+1})$, for $k =
0,..., p-1$.

This proof will extend to higher forms, of degree $p = d/2$, the same
singleton behaviour of $4d$ electrodynamics when extended to the $AdS_{d+1}$
bulk.

Let $V^k$ be the Harish-Chandra module with highest weight (lowest energy)
$(k,w_k)$, let $v_0$ be a highest
weight vector in $V^k$, and let
$V_0^k$ be the subspace generated by the subalgebra $so(2p)$ from $v_0$.
This subspace can be identified with the space
$Z^k \otimes {\bf C}v_0$, where $Z^k$ is level $k$ in the Grassmann algebra
generated by
$z_1,...z_{2p}$. Let $\psi$ denote a vector in this space. Apply the energy
raising operators to obtain the vectors
$$
\bigl\{L_{+i}\psi\bigr\}_{\psi \in Z^k \otimes {\bf C}v_0, i = 1,...,2p}.
$$
There is an action of $so(2p)$ on this space. Project out the irreducible
component with highest weight $w_{k+1}$, spanned
by the vectors
$$
 (L_{+i} - L_{+j}z_i\partial_j)\psi.\eqno(A.1)
$$
Now apply the energy lowering operators to these vectors, using the
commutation relations
$$
[L_{-m},L_{+i}] = L_{im} + k\delta_{im},~~ [L_{ij},z_m] = \delta_{jm}z_i -
\delta_{im}z_j.
$$
The result
$$
L_{-m}(L_{+i} - L_{+j}z_i\partial_j)\psi =  0,~~ i,m = 1,...,2p,
$$
is immediate. This shows that the Harish-Chandra module $V^k$ contains
$V^{k+1}$ as an invariant subspace, and that there is a
unique equivariant map from one to the other, that sends the vectors (A.1)
in $V^k$ to the highest level in $V^{k+1}$.

Next, consider the representations $D(p+k, w_{p-k}), k = 0,...,p$. Let
$U^k$ be the Harish-Chandra module with highest weight $(p+k,
w_{p-k}$. Apply the energy raising operators to $\psi \in U_0^k = Z^{p-k}
\oplus {\bf C}u_0$, $u_0$ the highest
weigp-ht vector in  $U^k$. Again $SO(2p)$ acts on this space. The subspace
that contains the component with the highest weight
$w_{k-1}$ consists of the vectors
$$
\sum_{i=1}^{2p} L_+^i z_i\tilde \psi,~~ \tilde \psi \in Z^{p-k-1} \otimes
{\bf C}u_0.
$$
Apply the lowering operators, using the commutation relations as above, to get
$$
L_{-j}\sum L_+^i z_i \tilde \psi = \bigl(L_{ij} +
\delta_{ij}(p+k)\bigr)\tilde\psi = 0,~~ j = 1,...,2p.
$$
This shows that there is an extension $D(p+k,w_{p-k}) \rightarrow
D(p+k+1,w_{p-k-1})$. We conclude that all the representations
$$
D(p\pm k,w_{p-k}),~~ k = 0,...,p,
$$
are zero-center modules.
$$
\hskip-1.8cm E_0\uparrow
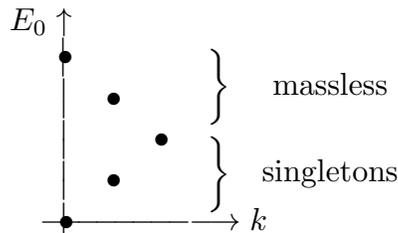
$$
\vskip-.7cm
$$
\eqalign{
& \hskip-2mm\matrix{\big|\hskip-1.2mm\bullet \cr \hskip-.3mm\big|&\bullet
\cr \hskip-.4mm\big|&&\bullet \cr \hskip-.3mm\bigl|&
\bullet
\cr
\hskip.2mm\big|\hskip-1.2mm\bullet
\cr}
\cr}
$$
\vskip-.9cm
$$
\hskip.7cm-\hskip-1.5mm -\hskip-1.5mm -\hskip-1.5mm -\hskip-1.5mm
-\hskip-1.5mm -\hskip-1.5mm -\hskip-1.5mm -\hskip-1.mm
-\hskip-1.5mm
\rightarrow k
$$
\vskip-2.8cm
$$
\hskip4cm \matrix{\biggl\}&{\rm massless}\cr  \cr\biggr\}&{\rm singletons}\cr}
$$
\bb
In the diagram, drawn for the case $p = 3$, we show the zero-center modules
\break
$D(p\pm k,w_{p-k})$ as dots at the points $E_0 = p\pm k$. We have shown
that there are extensions between neighbour
representations. In addition, we have seen that, as a field theory, the
singleton $D(2,w_1)$ carries the massless representation
$D(3,w_1)$ in its gauge subspace. We are confident that this is but an
example of a general phenomenon.

It is easy to show that all the extensions between neighbours are realized
on field modules. To do this we have only to generalize
the calculation of Section 2.3, replacing the formula (2.10) for the
(candidate) gauge modes by
$$
A = d\phi_1 + y\wedge \Phi_2 + y^2\Phi_3.
$$

\ve
\no{\bf Acknowledgements.}

S.F  is supported in part by DOE under grant DE-FG03-91ER40662, Task C, and
by ECC Science Program SCI*-CI92-0789 (INFN-Frascati)
and NSF Grant No. PHY94-07194. S.F. would like to thank the ITP Institute
in Santa Barbara for  kind hospitality during the
String-Duality workshop, where part of this work was done.

\bb

\no{\bf References.}

\item {[1]} C. Fronsdal, Phys. Rev. D{\bf 26} (1982) 1988; \hfill\break
I. Bars and M. Gunaydin, Commun. Math. Phys.
{\bf 87} (1982) 159; {\bf 91} (1983) 21; \hfill\break
M.P. Blencowe and M.J. Duff, Phys. Lett. {\bf B203} (1988)
229; \hfill\break
 H. Nicolai, E Sezgin and Y. Tanii, Nucl. Phys {\bf B305} (1988) 483;
\hfill\break
E. Bergshoeff, A. Salam, E. Sezgin and Y. Tanii, Nucl. Phys. B{\bf 305}
(1988) 497; Phys. Lett.
205 (1988) 237;  \hfill\break 
M. Gunaydin and D. Minic, hep-th/9802047; M. Gunaydin, hep-th/9803138.

\item {[2]} S. Ferrara, C. Fronsdal and A. Zaffaroni, hep-th/9802203, to
appear in Nucl.Phys.B.

\item {[3]} M. Gunaydin, L.J. Romans and N.P. Warner, Phys.Lett. {\bf 154B}
(1985) 268;\break
M.Gunaydin and N. Marcus,
Class. Quant. Gravity  {\bf 2} (1985) L11; \hfill\break
H.J. Kim, L.J. Romans and P. van Nieuwenhuizen, Phys Rev. {\bf D32} (1985) 389.

\item {[4]} M. Pernici, K.Pilch and P. van Nieuwenhuizen, Phys. Lett. {\bf}
(1984) 103.\hfill\break
M. Gunaydin, P. van Nieuwenhuizen and N.P.Warner, Nucl.Phys. {\bf B255}
(1985) 63.

\item {[5]} S. Ferrara and C. Fronsdal, hep-th/9712239, to appear in Class.
Quant. Grav.

\item {[6]} S. Ferrara and C. Fronsdal, hep-th/9802126, to appear in Phys.
Lett.

\item {[7]} M. Flato and C. Fronsdal, in {\it Essays in Supersymmetry}, (C.
Fronsdal Ed.), Kluwer 1986.

\item {[8]} B. Binegar, C. Fronsdal and W. Heidenreich, J. Mat, Phys. {\bf
24},2828 (1983).

\item {[9]} M. Flato and C. Fronsdal, J. Math. Phys. {\bf 22} 1100 (1981).

\item {[10]}  C. Fronsdal, Phys. Rev. {\bf D26} (1982) 1988.

\item {[11]} M. Flato and C. Fronsdal, Commun. Math. Phys. {\bf 108} (1987)
145.

\end